\documentclass[aps,prb,twocolumn,showpacs]{revtex4}

\usepackage{graphicx}
\usepackage{amsfonts}
\usepackage{amssymb}
\usepackage{amsmath}
\usepackage{mathrsfs}
\usepackage{braket}

\def\nn{\nonumber}

\def\ket#1{\vert #1 \rangle}

\newcommand{\kp}{k_{+}}
\newcommand{\ki}{k_{-}}
\newcommand{\idz}{i\partial_z}
\newcommand{\kperp}{\mathbf{k}_\parallel}

\begin{document}
\title{Microscopic simulation of superconductor-topological insulator proximity structures}
\author{Mahmoud Lababidi}
\affiliation{Department of Physics and Astronomy, George Mason University, Fairfax, Virginia, USA}
\author{Erhai Zhao}
\affiliation{Department of Physics and Astronomy, George Mason University, Fairfax, Virginia, USA}

\begin{abstract}
We present microscopic, self-consistent calculations of the 
superconducting order parameter and pairing correlations near the 
interface of an $s$-wave superconductor and
a three-dimensional topological insulator with spin-orbit coupling.
We discuss the suppression of the order parameter by the topological 
insulator and show that the equal-time pair correlation functions
in the triplet channel, induced by spin-flip scattering at the interface, 
are of $p_x\pm i p_y$ symmetry.
We verify that the spectrum at sub-gap energies is well described by
the Fu-Kane model. The sub-gap modes are viewed as interface states with
spectral weight penetrating well into the superconductor. 
We extract the phenomenological 
parameters of the Fu-Kane model from microscopic calculations, and find they
are strongly renormalized from the bulk material parameters. This is consistent 
with previous results of Stanescu et al for a lattice model using perturbation
theory in the tunneling limit.
\end{abstract}
\pacs{73.20.-r, 75.70.Tj, 85.75.-d}
\date{\today}
\maketitle

\section{Introduction}
Fu and Kane showed that 
at the interface between a three-dimensional topological band insulator  (TI) and an 
$s$-wave superconductor (S) forms a remarkable two-dimensional non-Abelian superconductor \cite{f-k}.
It hosts Majorana zero modes at vortex cores, as in a 
$p_x+ip_y$ superconductor \cite{r-g}, but respects time-reversal symmetry. 
As argued in Ref. \onlinecite{f-k}, the presence of superconductor induces a pairing interaction
between the helical Dirac fermions at the surface of the topological insulator,
and gaps out the surface spectrum. Then, the interface can be modeled elegantly by 
 a simple matrix Hamiltonian in Nambu space (we follow the convention of Ref. \onlinecite{qi-zhang}),
\begin{equation}
H_{FK}(\mathbf{k})=\left(
\begin{array}{cc}
h_s(\mathbf{k})  &  i\sigma_y  \Delta_s \\
-i\sigma_y \Delta_s^*  &   -h_s^*(-\mathbf{k})
\end{array}\label{fkmodel}
\right),
\end{equation}
where $\mathbf{k}=(k_x,k_y)$ is the two-dimensional momentum in
the interface plane, $\sigma_i$ are the Pauli matrices,
$h_s(\mathbf{k})$ is the surface Hamiltonian
for the topological insulator describing the helical Dirac fermions \cite{qi-zhang,rmp},
\begin{equation}
h_s(\mathbf{k})=-\mu_s + v_s (\sigma_xk_y-\sigma_y k_x).
\end{equation}
Fu and Kane also proposed to use S-TI proximity structures to
generate and manipulate Majorana fermions which obey non-Abelian 
statistics and are potentially useful for fault tolerant quantum 
computation \cite{f-k}. 
This proposal and a few others that followed based on 
superconductor-semiconductor heterostructures \cite{roman,maryland,jason,mao1,mao2} have revived
the interest in superconducting proximity effect involving insulating/semiconducting 
materials with spin-orbit coupling. More complex S-TI proximity structures
with ferromagnets \cite{yu,jacF} or unconventional superconductors \cite{jacU} 
have been investigated.

Given the importance of this proposal, it is desirable to
understand to what extent the effective model $H_{FK}$ holds,
and what are the values of $(\Delta_s,\mu_s,v_s)$ for given materials. 
Answering these questions is crucial for future experiments designed to probe
and manipulate Majorana fermions. As a first step in this
direction, Stanescu et al considered a microscopic lattice model
for the TI-S interface \cite{stan}. In this model, TI and S are described by a tight binding 
Hamiltonian defined on the diamond and hexagonal lattice respectively.
The two materials are coupled by tunneling term in the Hamiltonian. 
These authors found that 
for small $\mathbf{k}$, $H_{FK}(\mathbf{k})$ is valid but its parameters are 
significantly renormalized by the presence of the superconductor. 
This is supported by leading order 
perturbation theory in the weak coupling (tunneling) limit. They also
discussed the induced $p$-wave correlation within the framework of
perturbation theory. The $p$-wave correlation has also been noted in 
an analogous proximity structure in two dimension between
a quantum spin Hall insulator and a superconductor \cite{ann}.

In this work, we consider S-TI proximity structures where 
S and TI are {\it strongly} coupled to each other, 
rather than being separated by a tunneling barrier.
This is the desired, presumably the optimal, configuration to realize the Fu-Kane
proposal, e.g. to achieve maximum value of $\Delta_s$ in $H_{FK}$ for given
superconductor. 
In the strong coupling limit, 
the modification of superconductivity by the TI becomes important.
This includes the suppression of the superconducting order parameter,
the induction of triplet pair correlations by spin-active scattering
at the interface, and the formation of interface states below the bulk superconducting gap.
In order to accurately answer questions raised in the preceding paragraph
for strongly coupled S-TI structures, one has to self-consistently 
determine the the spatial profile of 
the order parameter near the interface.

Our work is also motivated by recent experimental discovery that Copper-doped 
topological insulator Cu$_x$Bi$_2$Se$_3$
becomes superconducting at a few Kelvins \cite{cu1,cu2}. It 
seems possible then to combine such superconductors with topological insulator 
Bi$_2$Se$_3$ to achieve strong proximity coupling. 
Using Bi$_2$Se$_3$ and Cu$_x$Bi$_2$Se$_3$ as one of the examples,
we set up microscopic, continuum models for the S-TI structures and solve the result 
Bogoliubov-de Gennes (BdG) equation numerically.
We first compute the superconducting order 
parameter as a function of the distance
away from the interface.
We then verify the validity of the Fu-Kane effective model
and extract its parameters from the low energy sector of the energy
spectrum. The emergence
of $H_{FK}$ will be viewed as the result of the ``inverse proximity effect", namely
strong modification of superconductivity by the presence of TI.
This is in contrast to the previous viewpoint of
pairing between surface Dirac fermions, which is a more proper description
in the tunneling limit.
The spectral weight of these low energy modes (with energy below the bulk superconducting gap) 
are shown explicitly to peak near the interface but penetrate well
into the superconductor.
We will also show analytically that the induced triplet pair
correlations are of $p_x\pm ip_y$ orbital symmetry, and systematically
study their spatial and momentum dependence.
Our results connect the phenomenological theory of Fu and Kane \cite{f-k}
to real materials. Our results for continuum models and strong coupling 
limit are also complementary to the results of Stanescu et al \cite{stan}
for lattice models and tunneling limit.

In what follows, we first outline the formulation of the problem and 
then present the main results. Technical details on numerically solving
the BdG equation are relegated to the appendix.

\section{Model and Basic equations}

The band gaps of topological insulators are much larger than the superconducting
gap of all  weak coupling $s$-wave superconductors. 
For the purpose of studying the proximity
effect between such superconductors and topological insulators, it is
sufficient to describe the topological insulator using the 
low energy effective $\mathbf{k}\cdot \mathbf{p}$ Hamiltonian. Following
Zhang et al \cite{zhang}, we model Bi$_2$Se$_3$ by 
\begin{eqnarray}
H_{TI}({\bf k})=
\left( \begin{array}{cccc}
M({\bf k}) & 0 & A_1 k_z  & A_2 k_{-} \\
0 &M({\bf k})& A_2 k_+ & -A_1 k_z \\
 A_1 k_z  & A_2 k_{-} &-M({\bf k}) & 0 \\
A_2 k_+ & -A_1 k_z &0 & -M({\bf k}) \\
 \end{array} \right)-\mu \hat{I}.
 \end{eqnarray}
Here $k_{\pm}=k_x \pm i k_y $, 
$M({\bf k})=M-B_1 k_z ^2 - B_2 (k_x^2 + k_y^2)$, and $\hat{I}$ is $4\times 4$
unit matrix. The numerical values of 
the parameters are obtained from first principle calculations \cite{zhang,band}, $M=0.28$ eV,
$A_1=2.2$ eV\AA, $A_2=4.1$ eV\AA, $B_1=10$ eV\AA$^2$, $B_2=56.6 $ eV\AA$^2$.
We work in basis $\{\ket{1\uparrow}, \ket{1\downarrow},\ket{2\uparrow},\ket{2\downarrow} \}$,
where 1 (2) labels the $P1_z^+$ ($P2_z^+$) orbital \cite{zhang}. Note that
we have neglected the unimportant diagonal term $\epsilon_0({\bf k})$ in Ref. \onlinecite{zhang} which 
only slightly modifies the overall curvature of the band dispersion. We also
keep the chemical potential $\mu$ as a tuning parameter.

\begin{figure}
\includegraphics[width=3in]{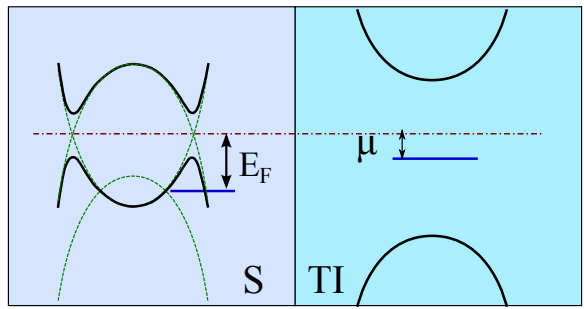}
\caption{Schematic (not to scale) band diagrams  in a superconductor-topological insulator (S-TI) proximity
structure. $E_f$ is the Fermi energy of the metal described by $H_M$ measured from the band crossing point. $\mu$ is
the chemical potential of TI measured from the band gap center. The superconducting gap
is much smaller than the band gap of TI.
}\label{setup}
\end{figure}

We consider a simple model of superconductor derived from a metallic
state obtained by turning off the spin-orbit coupling ($A_1=A_2=0$) in
$H_{TI}$ and tuning the Fermi level well into the conduction band \cite{zhao}. The metal
Hamiltonian
\begin{eqnarray}
H_{M}({\bf k})=\mathrm{diag} [M({\bf k}), M({\bf k}), -M({\bf k}), -M({\bf k})]-E_f \hat{I},
\end{eqnarray}
with  $E_f>M$. 
This mimics electron-doping the topological insulator \cite{cu2} 
or equivalently electrochemically
shifting its chemical potential by applying a gate voltage \cite{gate}.
As shown in Fig. \ref{setup}, 
the valence band (band 1 with dispersion $M({\bf k})-E_f$) is well below the Fermi level and remains inert as far as 
superconductivity is concerned.  Next, within the framework of Bardeen-Cooper-Schrieffer
theory, we assume attractive interaction between
the electrons in the conduction band (band 2) near the Fermi surface described by the reduced
 Hamiltonian,
\begin{equation}
H_{int}=\sum_{\bf k} \psi^\dagger_{2\uparrow}({\bf k})\psi^\dagger_{2\downarrow}(-{\bf k})\Delta + h.c.
\end{equation}
Here $\Delta$ is the 
superconducting order parameter, $\psi^\dagger_{l\sigma}$ is the electron creation operator for orbital $\l=1,2$ and
spin $\sigma=\uparrow,\downarrow$. The superconductor is then described by 
\begin{equation}
H_S=\sum_{{\bf k},l,\sigma}\psi^\dagger_{l\sigma}({\bf k})H_M({\bf k})_{l\sigma,l\sigma}\psi_{l\sigma}({\bf k})+H_{int}.
\end{equation}
Note that $H_S$ and $H_{TI}$ are in the same 
basis.

This model is inspired partly by superconductor Cu$_x$Bi$_2$Se$_3$.
The transition temperature at optimal doping $x=0.12$ is $T_c=3.8$K, which corresponds to a zero 
temperature superconducting gap $\Delta\sim$0.6meV \cite{cu1,cu2}. The Fermi level is 0.25eV 
above the bottom of the conduction band, and the Fermi wave vector $k_f\sim 0.12$\AA$^{-1}$.
The pairing symmetry of Cu$_x$Bi$_2$Se$_3$ is to our best knowledge not clear at present. 
If it turns out to be a conventional $s$-wave superconductor, its mains 
features will be captured by $H_S$ above with suitable choice of $E_f$ and $\Delta$.

%%%%%%%%%%%%%%%%%%%%%%%%%%
%%%%%%%%%%%%%%%%%%%%%%%%%%

Now consider a proximity structure consisting 
of a superconductor at $z<d$ and a topological insulator at $z>d$ (Fig. \ref{setup}).
The interface at $z=d$ is assumed to be specular, so the momentum $\mathbf{k_\parallel}=(k_x,k_y)$ 
parallel to the interface is conserved. The Hamiltonian for the whole system 
\begin{align}
\mathcal{H} = &\int d \mathbf{k}_\parallel dz
\Bigl\{ \sum_{\sigma} \psi^{\dagger}_{1 \sigma}(\mathbf{k}_\parallel,z)[h_0 - \mu(z)] \psi^{\dagger}_{1 \sigma}(\mathbf{k}_\parallel,z) 
 \nonumber \\
&- \sum_{\sigma} \psi^{\dagger}_{2 \sigma}(\mathbf{k}_\parallel,z)[h_0 + \mu(z)] \psi^{\dagger}_{2 \sigma}(\mathbf{k}_\parallel,z)\nonumber \\
&+\Delta(z)  \psi^{\dagger}_{2 \uparrow}(\mathbf{k}_\parallel,z) \psi^{\dagger}_{2 \downarrow} (-\mathbf{k}_\parallel,z)+h.c.\nonumber \\
&+A_1(z) [\psi^{\dagger}_{1 \uparrow} (-\idz) \psi_{2 \uparrow} 
+\psi^{\dagger}_{1 \downarrow} (\idz) \psi_{2 \downarrow}+h.c.]  \nonumber\\
&+A_2(z)
[\psi^{\dagger}_{1 \uparrow} \ki \psi_{2 \downarrow} +\psi^{\dagger}_{1 \downarrow} \kp \psi_{2 \uparrow} +h.c.]
\Bigr \}. \label{hbH}
\end{align}
Here $h_0(\mathbf{k}_\parallel, \partial_z)=M -B_1 \partial_z^2 -B_2 k_\parallel^2$, $\mu(z)$ and $A_i(z)$
are piece-wise constant, 
\begin{align}
\mu(z)&=E_f\theta(d-z)+\mu\theta(z-d),\\
A_i(z)&=A_i\theta(z-d),\;\;i=1,2
\end{align}
in terms of the step function $\theta$. The order parameter obeys the gap equation
\begin{equation}
\Delta(z) = g(z) \int d \mathbf{k}_\parallel \langle {\psi_{2 \uparrow}(\mathbf{k}_\parallel,z) \psi_{2 \downarrow}(-\mathbf{k}_\parallel,z) }\rangle. \label{gapE}
\end{equation}
We assume $g(z)=g\theta(d-z)$, the coupling constant $g$ determines the bulk gap.

To self-consistently solve Eq. \eqref{hbH} and \eqref{gapE}, we introduce
Bogoliubov transformation
\begin{equation}
 \psi_{l \sigma}(\mathbf{k}_\parallel,z) = \sum_n  u_{n,l \sigma} (\kperp, z) \gamma_{n,\kperp} + v^\ast_{n,l \sigma} (\kperp, z) \gamma_{n,\kperp}^\dagger
\end{equation}
to diagonalize $\mathcal{H}$ as
\begin{equation}
\mathcal{H} = E_g + \int d \kperp\sum_n \epsilon_n(k_\parallel) \gamma_{n,\kperp}^\dagger \gamma_{n,\kperp} ,
\end{equation}
where $E_g$ is the ground state energy, and $\gamma_{n,\kperp}^\dagger$
is the creation operator of Bogoliubov quasiparticles with energy $\epsilon_n(k_\parallel)$.
The wave function $u$ and $v$ satisfy the following Bogliubov-de Gennes (BdG) equation,
\begin{equation}
\hat{H}_{B}(\mathbf{k}_\parallel,z) \hat{\phi}_n(\mathbf{k}_\parallel,z)=\epsilon_n(k_\parallel)\hat{\phi}_n (\mathbf{k}_\parallel,z).
\label{bdgsimp}
\end{equation}
Here, the BdG Hamiltonian
\begin{equation}
\hat{H}_{B}=\left( \begin{array}{cccc}
h_0 -\mu  & \mathbf{d} \cdot \boldsymbol{\sigma}&0&0\\ 
 \mathbf{d} \cdot \boldsymbol{\sigma} &-h_0 -\mu &0&-\Delta\, i\sigma_y\\ 
0 &0& \mu -h_0 & \mathbf{d} \cdot \boldsymbol{\sigma}^* \\
0 &\Delta^{\ast} \, i\sigma_y & \mathbf{d} \cdot \boldsymbol{\sigma}^* & \mu+h_0\\ 
 \end{array} \right), \label{bdgH} 
\end{equation}
and the wave function (dropping the arguments)
\begin{equation}
\hat{\phi}_n=(u_{n,1\uparrow}, u_{n,1\downarrow}, u_{n,2\uparrow}, u_{n,2\downarrow}, 
v_{n,1\uparrow}, v_{n,1\downarrow}, v_{n,2\uparrow}, v_{n,2\downarrow})^\mathrm{T}.
\end{equation}
The vector $\mathbf{d}(\mathbf{k}_\parallel,z)$ is defined as
\begin{equation}
d_x=A_1(z)k_x,\,\, d_y=A_1(z)k_y,\,\, d_z=A_2(z)(-i\partial_z).
\end{equation} 
Other quantities such as $h_0(\mathbf{k}_\parallel,z)$, $\mu(z)$, and $\Delta(z)$ are 
defined above.
In terms of the wave functions, the zero temperature gap equation becomes
\begin{equation}
\Delta(z) = g(z) \int d \mathbf{k}_\parallel \sum_n' u_{n,2 \uparrow}(\mathbf{k}_\parallel,z) v^*_{n,2\downarrow}(-\mathbf{k}_\parallel,z) ,
\end{equation}
where the summation denoted by prime is restricted to $0<\epsilon_n<\omega_D$ with 
$\omega_D$ being the Debye frequency.

We will exploit a particular symmetry of the BdG Hamiltonian to simplify 
calculations. Define the polar angle $\varphi_k$ for the in-plane wave vector $\kperp$, 
\begin{equation}
k_x+ik_y=k_{\parallel}e^{i\varphi_k}.
\end{equation}
Then the BdG Hamiltonian for arbitrary $(k_x,k_y)$ is
related to that for $(k_x=k_\parallel,k_y=0)$ by unitary transformation 
\begin{equation}
\hat{U}^\dagger(\kperp) \hat{H}_{B}(k_x,k_y) \hat{U}(\kperp) = \hat{H}_{B}(k_\parallel,0).
\label{symm}
\end{equation}
Here $U$ is a 
block diagonal matrix,
\begin{equation}
U(\kperp)=\mathrm{diag}[e^{-i\sigma_z\frac{\varphi_k}{2}}, e^{-i\sigma_z\frac{\varphi_k}{2}}, e^{i\sigma_z\frac{\varphi_k}{2}},e^{i\sigma_z\frac{\varphi_k}{2}} ]. \label{unit}
\end{equation}
Thus, the eigen energy $\epsilon_n$ only depends on the magnitude of $\kperp$.
Once the wave function for $\varphi_k=0$ is known, the wave function for $\varphi_k\in (0,2\pi)$
can be obtained by simple unitary transformation.

We solve the matrix differential equation \eqref{bdgsimp} by conserving it into an algebraic equation, 
following the treatment of superconductor-ferromagnet structure by 
Halterman and Valls \cite{h-v}. The whole S-TI proximity structure is assumed to have 
finite dimension $L$ in the $z$
direction. The superconductor occupies the region $0<z<d$,
while the topological insulator occupies $d<z<L$. Hard wall boundary conditions are enforced at the end points, 
$z=0$ and $z=L$.  
The exact boundary conditions at the end points only affect the local physics there, provided 
that the boundaries are sufficiently far away from the S-TI interface. We expand the wave function
and order parameter in Fourier series \cite{h-v},
\begin{align}
u_{n,l\sigma}(z) &= \sum_m u_{nm}^{l \sigma}\,\phi_m(z),\label{uexp}\\ 
v_{n,l\sigma}(z) &= \sum_m v_{nm}^{l \sigma}\, \phi_m(z),\\
\quad \Delta(z) &= \sum_m \Delta_{m}\, \phi_m(z) , \label{Dexp}\\
\phi_m(z)&=\sqrt{2/L}\sin(k_m z).
\end{align}
The integer $m=1,2,...,N$ labels the quantized longitudinal (along $z$) momentum $k_m=m\pi/L$. 
The cutoff $N$ is chosen as \cite{s-v}
\begin{equation}
B_1k^2_N=M+E_f+\omega_D. \label{eq-N}
\end{equation}
By expansion Eq. \eqref{uexp}-\eqref{Dexp}, the BdG equation
becomes an $8N \times 8N$ matrix equation. With a reasonable guess of the order parameter profile, 
the eigen energies and eigen wave functions are obtained by solving the matrix eigen value problem.
Then a new order parameter profile is computed from the gap equation. The procedure is iterated
until convergence is achieved.  Relevant technical details can be found
in the appendix. 

To analyze the spectrum of the system, it is convenient to define the retarded Green's function
\begin{equation}
G^R_{l\sigma}(\mathbf{k}_\parallel,z,t)=-i\theta(t)\langle \{\psi_{l\sigma}(\mathbf{k}_\parallel,z,t),
\psi^\dagger_{l\sigma}(\mathbf{k}_\parallel,z,0)\}\rangle
\end{equation}
where the time-dependent field operators are in Heisenberg picture. 
For given $\kperp$ and $z$, the spectral functions 
are defined as
\begin{align}
&N_{l\sigma}(\mathbf{k}_\parallel,z,\omega)= -\mathrm{Im}G^R_{l\sigma}(\mathbf{k}_\parallel,z,\omega), \\
&N(\mathbf{k}_\parallel,z,\omega)=\sum_{l\sigma}N_{l\sigma}(\mathbf{k}_\parallel,z,\omega).
\end{align}
In terms of the wave functions and eigen energies, 
\begin{equation}
N_{l\sigma}(\mathbf{k}_\parallel,z,\omega>0)=\sum_n|u_{n,l\sigma}(\mathbf{k}_\parallel,z)|^2\delta(\omega-\epsilon_n). 
\end{equation}
We also introduce the equal-time pair correlation functions
for the conduction electrons 
\begin{equation}
F_{\alpha\beta}(\mathbf{k}_\parallel,z)=\langle \psi_{2\alpha}(\mathbf{k}_\parallel,z) \psi_{2\beta}(-\mathbf{k}_\parallel,z)\rangle.\label{pair-corr}
\end{equation}
For example, at zero temperature we have
\begin{align}
F_{\uparrow\uparrow}(\mathbf{k}_\parallel,z)=\sum'_n u_{n,2\uparrow}(\mathbf{k}_\parallel,z)
v^*_{n,2\uparrow}(-\mathbf{k}_\parallel,z),\\
F_{\downarrow\downarrow}(\mathbf{k}_\parallel,z)=\sum'_n u_{n,2\downarrow}(\mathbf{k}_\parallel,z)
v^*_{n,2\downarrow}(-\mathbf{k}_\parallel,z).
\end{align}
Triplet components of $F$ will be induced near the S-TI interface by spin-active
scattering \cite{zhao}.

\section{the order parameter}

%%%%%%%%%%%%%%%%%%%%%%%%%%
\begin{figure}
\includegraphics[width=3.4in]{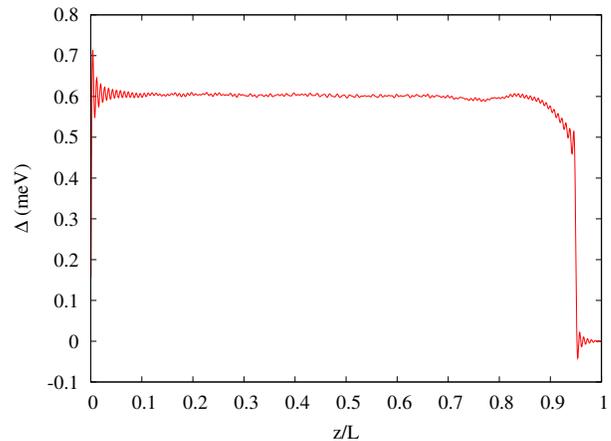}
\caption{The superconducting order parameter $\Delta(z)$ near an S-TI interface at $
z=d=0.95L$. The superconductor occupies $0<z<d$, and topological insulator occupies $d<z<L$. 
$L=300$ nm, $\mu$=0, the bulk gap $\Delta_0=$0.6meV. }\label{delta-cu}
\end{figure}
%%%%%%%%%%%%%%%%%%%%%%%%%%

First we present the spatial profile of the superconducting order parameter $\Delta(z)$
after the convergence is achieved. In all following calculations, $E_f$ is fixed at 0.4eV, 
which is modeled after optimally doped Cu$_x$Bi$_2$Se$_3$ \cite{cu2}. And the Debye frequency
is set as $\omega_D=0.1E_f$ \cite{h-v}.
Fig. \ref{delta-cu} shows an example with $\mu=0$, $L=300$nm, $d=0.95L$, and
a bulk gap of 0.6meV as found in Cu$_x$Bi$_2$Se$_3$. 
Going from the superconductor into the topological insulator,
$\Delta$ first gets suppressed as the interface is approached before it
drops to zero inside TI. The suppression is roughly 20\% at the interface.
Note that the fine wiggles of $\Delta$ in the simulation results 
are due to the finite momentum cutoff of 
the longitudinal momentum $k_m$. As previously discussed by 
Stojkovic and Valls \cite{s-v}, the number of oscillations is $\sim N/2$, and the 
oscillation amplitude vanishes in the bulk as $N$ is increased. 
In this case, $N$ is chosen to be
258 according to Eq. \eqref{eq-N}. So the matrix to be diagonalized is 2064 by 2064.

\begin{figure}
\includegraphics[width=3.4in]{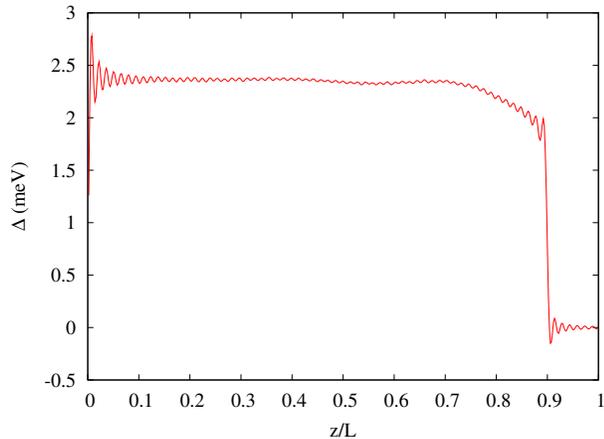}
\caption{The order parameter $\Delta(z)$ near an S-TI interface at $
z=d=0.9L$. $L=160$ nm, $\mu$=0, $\Delta_0\sim 2.4$meV.}
\label{delta-24}
\end{figure}

\begin{figure}
\includegraphics[width=3.4in]{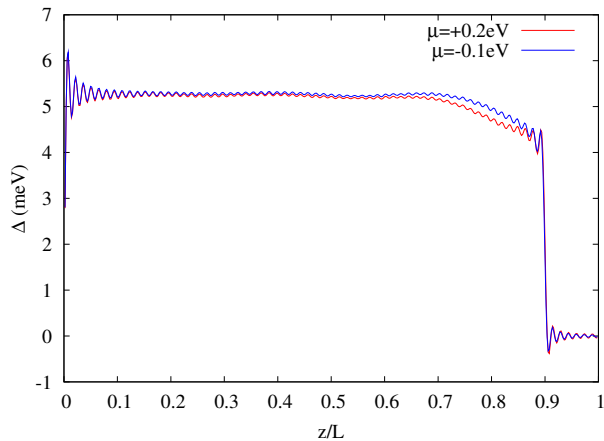}
\caption{The order parameter profile for two different chemical potentials of
the topological insulator, $\mu=-0.1$eV and $\mu=0.2$eV. $L=160$nm,  
$\Delta_0\sim 5.2$meV. }\label{delta-chem}
\end{figure}

Fig. \ref{delta-24} show the result for $\mu=0$, $d=0.9L$, and 
a superconductor with bulk gap $\Delta_0\sim 2.4$meV. Since the coherence length
is much smaller than the previous example, it is sufficient to consider 
$L=160$nm, and correspondingly $N=138$. 
The order parameter profile  
depends weakly on $\mu$, as shown in Fig. \ref{delta-chem} for a superconductor
with bulk gap $\sim 5.2$meV. 
From these examples, one observes that the length scale over which $\Delta$ is 
significantly suppressed does {\it not} scale with $\xi_0$, the zero temperature
coherence length of the superconductor. Rather it stays roughly the same, 
on the order of $30$nm, as $\xi_0$ is varied over one decade from Fig. \ref{delta-cu}
to Fig. \ref{delta-chem} (note the horizontal axis is $z/L$). 
This is not very surprising since $\xi_0$ is not the only length scale at play here.
The interface represents a strong (as compared to $\Delta_0$) perturbation that 
significantly distorts the bulk wave functions.
The self-consistent microscopic BdG approach provides a reliable way to
capture the details of $\Delta(z)$ near the interface.

%%%%%%%%%%%%%%%%%%%%%%%%%%
%%%%%%%%%%%%%%%%%%%%%%%%%%

%%%%%%%%%%%%%%%%%%%%%%%%%%
%%%%%%%%%%%%%%%%%%%%%%%%%%

It is illuminating to compare the proximity effect in S-TI
structure with that in S-F structure \cite{toku}, where F stands for 
a ferromagnetic insulator. 
The presence of F breaks time-reversal and spin rotation symmetry and 
significantly suppresses the order parameter. The suppression is
sensitive to the spin mixing angle which is related to the band gap
and exchange field of F \cite{toku}.
In contrast, despite the spin-active scattering of electrons 
by TI which introduces spin-flips and spin-dependent phase shifts \cite{zhao}, 
spin-orbit coupling is not pair breaking.
The suppression of $\Delta$ near the interface is to a large extent 
due to the reorganization of local wave functions enforced by the boundary conditions 
at $z=d$ for piece-wise potentials $\mu(z)$, $A_i(z)$, $g(z)$. 
It depends on for example how the wave functions decay inside the TI
for given $E_f$ and $\mu$, and involves ``high-energy" physics beyond the
scale of $\Delta$ but below the scale of the band gap. 
To test this, we have investigated the proximity effect between the same
superconductor and
a hypothetical ordinary insulator modeled by $H_{TI}$ with $A_1=A_2=0$
and the same band gap. The suppression of $\Delta$ by such an 
ordinary insulator turns out to be very similar. 

\section{The interface mode and the Fu-Kane model}

\begin{figure}
\includegraphics[width=3.4in]{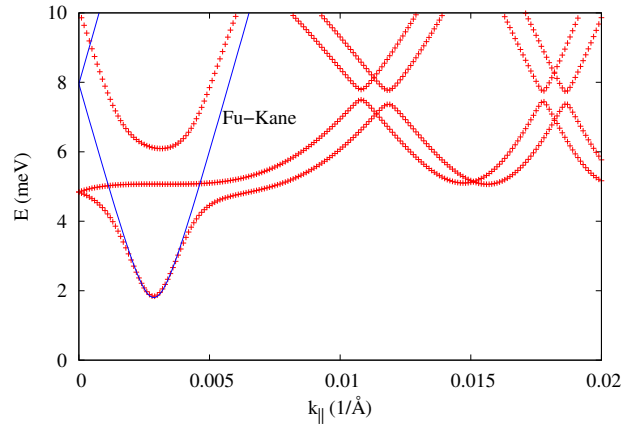}
\caption{The lowest few energy levels $\epsilon_n(k_\parallel)$. 
$\mu=0$, $L=160$nm, and the bulk superconducting gap $\Delta_0\sim$5.2meV.
A well-defined interface mode is clearly visible at sub-gap energies.
Solid lines show a fit to the Fu-Kane model, with $\Delta_s=1.8$meV, 
$v_s=2.7$eV\AA, and $\mu_s=7.5$meV.
}\label{lev-fk}
\end{figure}

Next we analyze the energy spectrum of the system, $\epsilon_n(k_\parallel)$,
obtained from the BdG calculation. 
Take the case of $\mu=0$, $L=160$nm, $d=0.9L$, $\Delta_0\sim 5.2$meV as an example.
Fig. \ref{lev-fk} shows the first several energy levels of the composite
system versus the transverse momentum $k_\parallel$. There are many continuously
dispersing modes at energies above the bulk gap. They are the usual Bogoliubov
quasiparticles for different quantized longitudinal momenta.
One also sees a series of avoided level crossings.
At small $k_\parallel$ emerges a well-defined mode below $\Delta_0$. We will 
identify it as the interface mode first discussed by Fu and Kane \cite{f-k}.

The Fu-Kane model Eq. \eqref{fkmodel} predicts the dispersion 
\begin{equation}
E(k)=\sqrt{|\Delta_s|^2+(v_sk \pm\mu_s)^2}.
\end{equation}
We fit the very low energy portion of the spectrum to this prediction to
extract the phenomenological parameters in the Fu-Kane model. The result
is shown in Fig. \ref{lev-fk}. We find that,
not surprisingly, $\Delta_s=1.8$meV which is much smaller than $\Delta_0=5.2$meV, and 
$v_s=2.7$eV$\mathrm{\AA}$ which deviates significantly 
from $A_2=4.2$eV$\mathrm{\AA}$ predicted for the surface 
dispersion of TI. Moreover, $\mu_s=7.5$meV despite that the chemical potential
of TI is $\mu=0$. Therefore, our results show that the values of $(\Delta_s,v_s,\mu_s)$
are strongly renormalized by the presence of the superconductor. This is consistent 
with the findings of Stanescu et al for weakly coupled S-TI structures \cite{stan}. 

%%%%%%%%%%%%%%%%%%%%%%%%%%
%%%%%%%%%%%%%%%%%%%%%%%%%%

\begin{figure}
\includegraphics[width=3.4in]{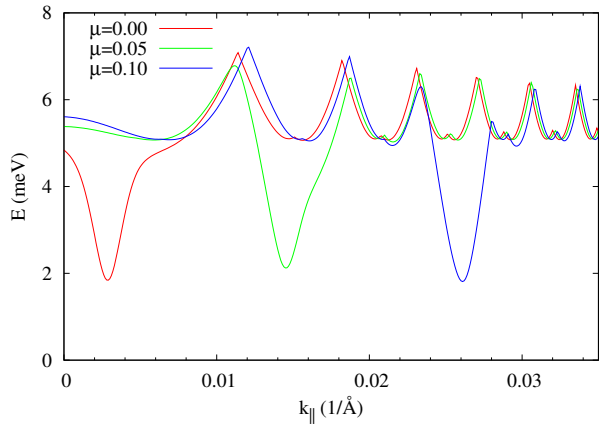}
\caption{The dispersion of the lowest energy level
for different $\mu$ (in eV). Other parameters
are the same as in Fig. \ref{lev-fk}, $L=160$nm and $\Delta_0\sim$5.2meV. Fu-Kane model
well describes the lowest energy mode. As $\mu$ is increased, 
$\Delta_s$ and $v_s$ stay roughly the same, while $\mu_s$ scales
linearly with $\mu$.
}\label{lev-chem}
\end{figure}

\begin{figure}
\includegraphics[width=3.4in]{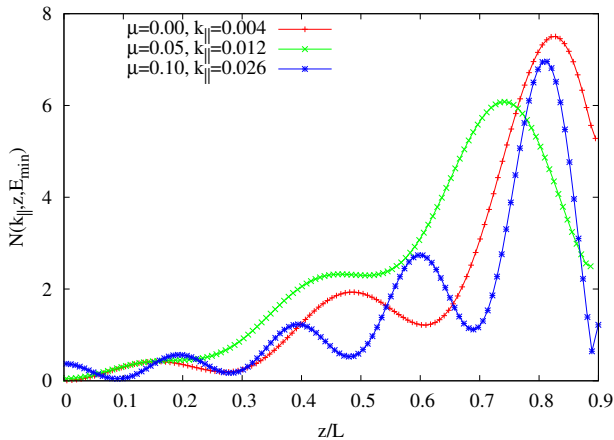}
\caption{The spectral function $N(k_\parallel,z,\omega)$ of the lowest 
energy level, $\omega=E_{min}$, shown in Fig. \ref{lev-chem}. 
The interface is at $z=0.9L$, $L$=160nm.
The spectral function oscillates rapidly with $z$, so only its envelope is plotted.
}\label{sp}
\end{figure}

We have checked the validity of the Fu-Kane model for a variety of chemical potentials.
Representative examples are plotted in Fig. \ref{lev-chem}. In each case, the sub-gap 
mode can be well accounted by the Fu-Kane 
model with suitable choice of parameters. While $\mu_s$ is always different from $\mu$,
numerically we find it scales linearly with $\mu$. At the same time, 
$\Delta_s$ and $v_s$ show no strong dependence on $\mu$ for this set of parameters.
To make sure that the sub-gap mode is indeed localized near the interface, we plot in 
Fig. \ref{sp} the $z$ dependence of the spectral function $N(k_\parallel,z,\omega)$.
The spectral weight of the sub-gap mode is peaked near the interface and decays over 
a length scale $\sim \xi_0$ into the superconductor.
Note that the spectral weight on the TI side (not shown in the figure) is finite, 
but it is much smaller in magnitude and decays very fast inside TI.

%%%%%%%%%%%%%%%%%%%%%%%%%%
%%%%%%%%%%%%%%%%%%%%%%%%%%

%%%%%%%%%%%%%%%%%%%%%%%%%%
%%%%%%%%%%%%%%%%%%%%%%%%%%

\begin{figure}
\includegraphics[width=3.4in]{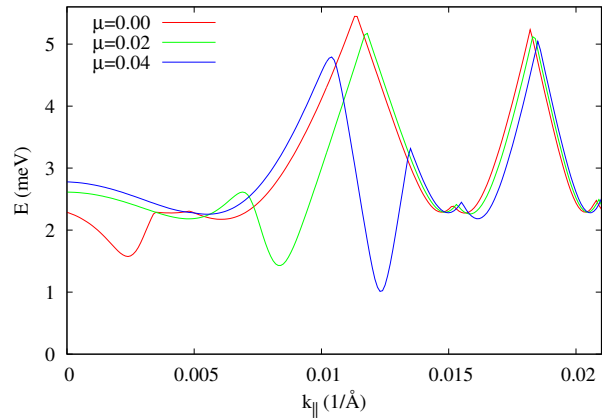}
\caption{The lowest energy level of an S-TI structure with $L=160$nm,
$d=0.9L$, $\Delta_0=2.4$meV. $\mu$ is the chemical potential of the TI and
measured in eV.}\label{level-27}
\end{figure}

We have carried out similar analysis for superconductors with larger
coherence length. Fig. \ref{level-27} shows the evolution of
the sub-gap mode with $\mu$ for $\Delta_0=2.4$meV. In this case,
the values of $(\Delta_s,v_s,\mu_s)$ all varies with $\mu$. 
Superconductors with larger $\xi_0$ and smaller $\Delta_0$ are thus more
sensitive to changes in $\mu$ and other microscopic details near
the interface. The exact values of the effective parameters 
in the Fu-Kane model in general depend on such microscopic details.

%%%%%%%%%%%%%%%%%%%%%%%%%%
%%%%%%%%%%%%%%%%%%%%%%%%%%

\section{Triplet pair correlations}

It is well known that in heterostructures of $s$-wave superconductors,
pairing correlations in other orbital channels, e.g. $p$-wave correlations, 
will be induced by scattering at the interfaces \cite{esch,tanaka}. For example,
inversion/reflection symmetry ($z\leftrightarrow -z$) is lost in an S-TI proximity 
structure, and the appearance of $p$-wave correlations seems natural from
 partial wave analysis. Moreover, scattering by a topological insulator is 
spin-active. The spin-orbit coupling inside a TI acts like a momentum-dependent
magnetic field to flip the electron spin and introduce different phase shifts
for spin up and down electrons. The scattering matrix has been worked out by us 
previously \cite{zhao}. Thus, a singlet $s$-wave Cooper pair can be converted into a pair
of electrons in spin-triplet state at the S-TI interface.
However, it is important to recall that by assumption attractive interaction only exists
(or is appreciable) in the $s$-wave channel. There is no binding force
to sustain a triplet Cooper pair or a triplet superconducting order parameter. 
Similar (but different) pairing correlations in superconductor-ferromagnet
hybrid structures have been extensively studied \cite{esch}. 
The appearance of $p$-wave correlations in S-TI systems
has been pointed out previously by Stanescu et al using a perturbative analysis \cite{stan}.

We focus on the equal-time pair correlation functions defined in Eq. \eqref{pair-corr}.
By exploiting the symmetry of the BdG Hamiltonian, Eq. \eqref{symm}, we are able to
find analytically the orbital structure of the triplet correlation functions. The unitary 
transformation Eq. \eqref{unit} yields
\begin{align}
u_{2\uparrow}(k_x,k_y)=u_{2\uparrow}(k_\parallel,0)e^{-i\varphi_k/2},\nn \\
u_{2\downarrow}(k_x,k_y)=u_{2\downarrow}(k_\parallel,0)e^{+i\varphi_k/2},\nn \\
v_{2\uparrow}(k_x,k_y)=v_{2\uparrow}(k_\parallel,0)e^{+i\varphi_k/2},\nn \\
v_{2\downarrow}(k_x,k_y)=v_{2\downarrow}(k_\parallel,0)e^{-i\varphi_k/2}. 
\end{align}
Using these relations, we find
\begin{align}
F_{\uparrow\uparrow}(\kperp,z)=F_{\uparrow\uparrow}(k_\parallel,z)e^{-i\varphi_k},\\
F_{\downarrow\downarrow}(\kperp,z)=F_{\downarrow\downarrow}(k_\parallel,z)e^{+i\varphi_k}.
\end{align}
Namely $F_{\uparrow\uparrow}$ ($F_{\downarrow\downarrow}$) has $p_x-ip_y$ ($p_x+ip_y$) orbital 
symmetry. Finally, the remaining triplet correlation function
\begin{equation}
\langle \psi_{2\uparrow}(\kperp,z) \psi_{2\downarrow} (-\kperp,z)+ \psi_{2\downarrow}(\kperp,z) \psi_{2\uparrow}(-\kperp,z)\rangle
\end{equation}
turns out to be zero. Note that the so-called odd-frequency paring correlations 
\cite{esch,tanaka,trip}, which vanishes in the equal-time limit, 
are also interesting in S-TI structures, but 
we will not discuss their behaviors here.

%%%%%%%%%%%%%%%%%%%%%%%%%%
\begin{figure}
\includegraphics[width=3.4in]{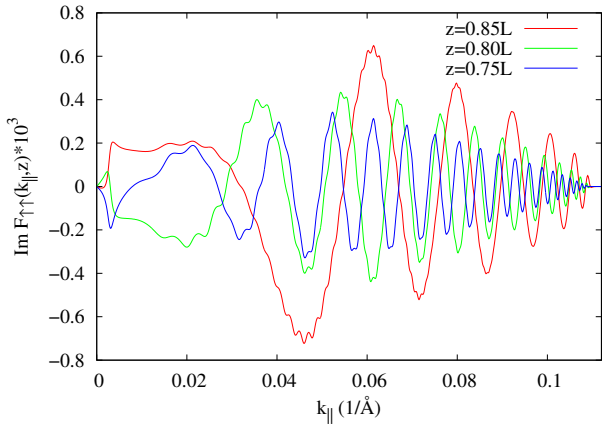}
\caption{The imaginary part of triplet pair correlation function 
$F_{\uparrow\uparrow}(k_\parallel,z)$. The S-TI interface is at
$d=0.9L$. $\mu=0$, $L=160$nm, $\Delta_0=5.2$meV.
}\label{pw}
\end{figure}
%%%%%%%%%%%%%%%%%%%%%%%%%%

\begin{figure}
\includegraphics[width=3.4in]{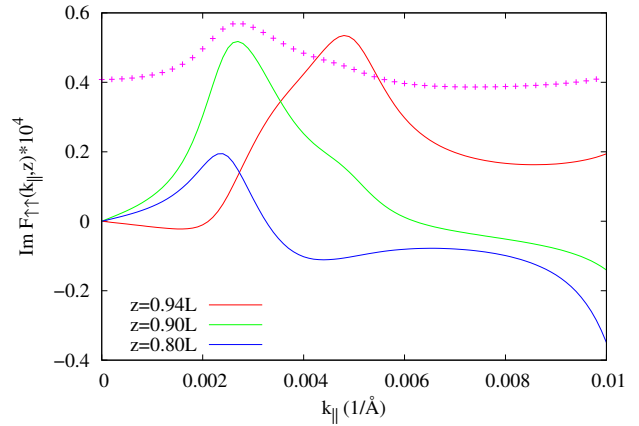}
\caption{The imaginary part of 
$F_{\uparrow\uparrow}(k_\parallel,z)$. $\mu=0$, $L=300$nm, $d=0.95L$, $\Delta_0=0.6$meV.
As comparison, the data points show the singlet pair correlation function 
$F_{\uparrow\downarrow}(k_\parallel,z=0.9L)/3$.
}\label{pw-cu}
\end{figure}

We find that $F_{\uparrow\uparrow}(k_\parallel,z)$ is 
purely imaginary and identical to $F_{\downarrow\downarrow}(k_\parallel,z)$.
The results for $\mu=0$, $L=160$nm, $d=0.9L$, $\Delta_0=5.2$meV are plotted
in Fig. \ref{pw}. $F_{\uparrow\uparrow}$ vanishes at $k_\parallel=0$
as well as for large $k_\parallel$, namely when $k_\parallel>\sqrt{(E_F+\omega_D+M)/B_2}$.
This is consistent with lack of pairing in both limits. 
The behavior of $F_{\uparrow\uparrow}$ for small $k_\parallel$
is illustrated in Fig. \ref{pw-cu} for $\mu=0$, $L=300$nm, $d=0.95L$, 
$\Delta_0=0.6$meV. As comparison, we also plotted the singlet
 pair correlation function 
\begin{equation}
F_{\uparrow\downarrow}(\mathbf{k}_\parallel,z)=\sum'_n u_{n,2\uparrow}(\mathbf{k}_\parallel,z)
v^*_{n,2\downarrow}(-\mathbf{k}_\parallel,z)
\end{equation}
which is $s$-wave and purely real.

%%%%%%%%%%%%%%%%%%%%%%%%%%
%%%%%%%%%%%%%%%%%%%%%%%%%%

\section{Summary and outlook}
In summary, we have investigated the proximity effect between an $s$-wave superconductor
and a topological insulator using a microscopic continuum model. 
Strong coupling between the two materials renders the surface state of TI a less 
useful concept for this problem.
Our focus has been on the various modifications to superconductivity by the presence of TI. 
These include the suppression of the order parameter, the formation of interface modes
below the bulk superconducting gap, and the induction of triplet pairing correlations.
It is gratifying to see the Fu-Kane effective model emerges in the low energy sector
albeit with a set of renormalized parameters. Our results are complementary to
previous theoretical work on the proximity effect \cite{f-k,stan} and confirm the validity
of the Fu-Kane model.

We made a few simplifying assumptions in our calculation. The superconductor is described 
by a two-band model with the valence band well below the Fermi level. Since only 
electrons near the Fermi surface are relevant for weak coupling superconductivity, we 
believe our main results are general. As idealizations, the chemical 
potential, the spin-orbit coupling, and the attractive interaction are assumed to be 
step functions with a sudden jump at the interface. 
More elaborate and realistic models can be considered within the framework of BdG equations.
For example, one can add a tunneling barrier between S and TI, 
or include a Rashba-type spin-orbit coupling term 
(due to the gradient of chemical potential) at the interface. We will not 
pursuit these generalizations here.
Finally, the approach outlined here can be straightforwardly applied to 
study non-Abelian superconductivity in other superconductor-semiconductor
heterostructures where spin-orbit coupling also plays a significant role
\cite{roman,maryland,jason,mao1,mao2}.

\section{acknowledgements}
This work is supported by NIST Grant No. 70NANB7H6138 Am 001 
and ONR Grant No. N00014-09-1-1025A. 

\appendix*
\section*{Appendix}
We follow the numerical scheme of Halterman and Valls to solve 
the matrix BdG equation \cite{h-v}. The wave functions and the order parameter
are expanded in the orthonormal basis $\{\phi_m(z)\}$, with $m=1,...,N$. For example,
function $u_{n,1\uparrow}(z)$ is represented by $N$ numbers,
\begin{align*}
(u^{1\uparrow}_{n,1},u^{1\uparrow}_{n,2},...u^{1\uparrow}_{n,m}...,u^{1\uparrow}_{n,N}).
\end{align*}
Accordingly, each term in $\hat{H}_{B}$ is represented
by a $N\times N$ matrix with the matrix elements given by
\begin{align*}
h_0(\kperp,\partial_z) &\rightarrow \delta_{mm'}(M - B_1 k_m^2 - B_2 k_\parallel^2) \\
U(z) &\rightarrow E_f E_{mm'} + \mu F_{mm'} \\
 A_2(z) \partial_z &\rightarrow A_2 G_{mm'}\\
 A_1(z) k_\pm &\rightarrow A_z k_\pm F_{mm'} \\
 \Delta &\rightarrow D_{mm'}\equiv \sum_{m''}J_{m,m',m''}\Delta_{m''}
\end{align*}
where 
\begin{align*}
E_{mm'}&=\int_0^{d} \phi_m(z) \, \phi_{m'}(z) dz \\
F_{mm'}&=\int_d^{L} \phi_m(z) \, \phi_{m'}(z) dz \\
G_{mm'}&=\int_d^{L} \phi_m(z) \partial_z \phi_{m'}(z) dz\\
J_{m,m',m''}&=\int_0^{d} \phi_m(z) \, \phi_{m'}(z) \, \phi_{m''}(z) dz
\end{align*}
These integrals can be evaluated analytically. Then the BdG equation becomes
an $8N\times 8N$ matrix equation. The gap equation can be rewritten as
\[
\Delta_{m}=g\int d\kperp \sum'_n \sum_{m',m''} J_{m,m',m''}
u^{2\uparrow}_{nm'}(\kperp)v^{2\downarrow}_{nm''}(-\kperp)^* 
\]
The integral over $\kperp$ is first simplified to an integral over $k_\parallel$
by the symmetry Eq. \eqref{symm} and then evaluated numerically with high 
momentum cutoff 
$\sqrt{(E_F+\omega_D+M)/B_2}$.

\bibliographystyle{apsrev}

\end{document}